\documentclass[12pt]{article}

\usepackage{bigstrut}
\usepackage{comment}
\usepackage[a4paper, total={6in, 8in}]{geometry}
\usepackage{mathrsfs}
\usepackage{authblk}
\usepackage[normalem]{ulem}
\usepackage{bm}
\usepackage[leftcaption]{sidecap} 

\title{\Large \textbf{Capturing multiscale interactions in fluid flow via Lagrangian coherent structures and modal analysis}}
\small
\author{\normalsize Morgan R. Jones$^{1,2}$}
\author{\normalsize Charles J. Klewicki$^{1,3}$}
\author{\normalsize Oliver Khan$^{1,4}$}
\author{\normalsize Steven L. Brunton$^5$}
\author{\newline\normalsize Mitul Luhar$^1$}
\date{}
\affil{\normalsize $^1$Department of Aerospace and Mechanical Engineering, \\University of Southern California, Los Angeles, CA 90010}
\affil{\normalsize $^2$Los Alamos National Laboratory, Los Alamos, NM 87544}
\affil{\normalsize $^3$Department of Mechanical and Aerospace Engineering, \\
University of California, Los Angeles, CA 90095}
\affil{\normalsize $^4$Department of Mechanical Engineering, \\Stanford University, Stanford, CA 94305}
\affil{$^5$Department of Mechanical Engineering, \\University of Washington, Seattle, WA 98195}

\usepackage{amsmath}
\usepackage{verbatim}
\usepackage{hyperref}
\usepackage{ragged2e}
\usepackage{setspace}
\DeclareMathOperator{\sech}{sech}

\newcommand{\ub}{\bm{u}}
\newcommand{\xb}{\bm{x}}

\usepackage[sc]{mathpazo}
\usepackage[T1]{fontenc}

\hypersetup{
    colorlinks=true,
    linkcolor=blue,
    filecolor=magenta,      
    urlcolor=blue,
    citecolor=blue,
}
\usepackage{xcolor}
\usepackage{caption}

\usepackage[numbers, sort&compress]{natbib}
\usepackage{graphicx}
\usepackage{float}
\graphicspath{ {./images/} }

\begin{document}
{\small \maketitle} 
%
\begin{abstract}
We consider the relationship between Eulerian modal decompositions and Lagrangian coherent structures (LCSs). The model sensitivity framework developed by \citet{kaszas_universal_2020} is used to express data-driven modal representations of fluid flow in a Lagrangian space. The method, based on the computation of the finite-time Lyapunov exponent, computes the amplitude perturbations experienced by fluid particles due to specific modal components of the flow. Demonstrations of the method are presented for both periodic and turbulent flows, including direct numerical simulation (DNS) data of the classical cylinder wake flow, experimental data from the wake past an oscillating foil, and DNS data of a turbulent channel flow. This method provides a way to understand how Eulerian mode structures interact dynamically with features of the Lagrangian coherent structure across scales, offering additional physical insight into modal decompositions.
\end{abstract}
\begin{keywords} 
\smallskip Lagrangian coherent structures, modal analysis, chaos, low-dimensional models
\end{keywords}

\section{Introduction}

Fluid flows are characterized by a range of complex processes such as instabilities, turbulent mixing, and vortex interactions. When these features are present, it is often useful to interpret and visualize fluid flows in terms of dominant patterns or \textit{coherent structures}. Two widely used approaches for understanding coherent structures are modal decompositions and Lagrangian coherent structure (LCS) analyses.

Modal decomposition methods have advanced over the past decade, capturing the coherent structures of interest in the form of modes. In practice, one may visualize contributions from individual modes (or small sets of modes) to better understand the dominant features of the flow field. Techniques such as proper orthogonal decomposition (POD) and dynamic mode decomposition (DMD) identify energetically or dynamically important modes that can be linearly superposed to reconstruct the flow field \citep{taira_modal_2017}. The linear properties of these decompositions are attractive from a mathematical point of view, and for interpreting modes as additive components of the full flow field. These algorithms, as well as extensions such as spectral POD, balanced POD, and many variants of DMD, have been used successfully to identify and represent flow patterns such as vortex shedding, vortex pairing and merging, Kelvin-Helmholtz instabilities, and very-large-scale motions (VLSMs) in turbulent flows \citep[e.g.,][]{taira_modal_2020,saxton-fox_amplitude_2022,Jones_Kanso_Luhar_2024}. 

In practice, mode structures are extracted from Eulerian measurements of flow variables such as velocity, vorticity, or pressure. However, it has been argued that the identification of \textit{instantaneous} flow features that are dynamically influential from such Eulerian analyses are generally frame-dependent and heuristic, which limits their reliability \citep{green_detection_2007,haller_lagrangian_2015}. Moreover, these representations often require a user-defined threshold for visualization, which can lead to difficulties in analyses such as vortex identification. Despite the limitations, modal analysis techniques have enabled significant progress towards the development of computationally efficient reduced-order models that provide physical insight and inform flow control \citep{brunton_closed-loop_2015,taira_modal_2020}. 

Fluid flow can alternatively be analyzed in terms of Lagrangian coherent structures (LCSs), which can be defined as material surfaces that attract, repel, or trap fluid parcels. A data-driven method for identifying LCSs is the finite-time Lyapunov exponent (FTLE) \citep[e.g.,][]{shadden_definition_2005,haller_lagrangian_2015}. FTLEs characterize regions of maximum strain in the flow with respect to particle initial conditions. Thus an FTLE field can identify material surfaces that govern particle trajectories in the form of stable manifolds where particles converge and unstable manifolds that push particles apart. For fluid flows, the FTLE field can be useful for characterizing mixing regions and identifying vortices in an objective frame \citep{green_unsteady_2011,green_detection_2007}. In a sense, LCSs are the hidden skeleton of fluid flow that organize fluid particle motion and transport \citep{haller_lagrangian_2015}.

Hybrid approaches that combine modal or hierarchical representations with LCSs offer a promising route to better understanding coherent structures. Although less explored in the fluid mechanics community, there have been prior efforts involving such hybrid analyses.
\citet{macmillan_lagrangian_2022} for instance demonstrated a scale decomposition that was Lagrangian and based on transport. This method uses a graph Fourier transform approach, whereby an eigendecomposition of the graph Laplacian identifies structures with specific spatial scales. \citet{xie_lagrangian_2020} used a Lagrangian inner product to combine Eulerian and FTLE data, leading to more accurate approximations of the FTLE field and streamfunction, with basis functions that differed in scale from those based purely on Eulerian data. These approaches, hold significant potential for revealing new coherent structures relevant to particle trajectories while also capturing multi-scale features embedded within the flow. 

This paper extends these concepts by studying how certain modes obtained from Eulerian measurements relate to the LCSs of a flow. In this context, we use the \textit{model sensitivity} framework developed by \citet{kaszas_universal_2020}. While the FTLE field characterizes the sensitivity of particle trajectories to initial conditions, the model sensitivity framework can be used to account for stochastic and deterministic perturbations to the dynamical system itself. In the context of modal decomposition, we use this framework to consider the contribution of specific mode structures to fluid particle motion and transport. To distinguish this approach, which aims to highlight the contribution of modal components of various scale, from the more general model sensitivity framework, we use the term \textit{modal-trajectory uncertainty}.

The following Section \ref{sect:msframework} provides a brief review of the model sensitivity framework developed by \citet{kaszas_universal_2020}. As an illustrative example, the framework is used to evaluate the influence of a cross-stream perturbation on the FTLE field for a simple kinematic model. Section \ref{sect:modal-trajectory uncertainty} builds on the model sensitivity framework to consider perturbations in the context of modal representations. The subsequent examples in Section \ref{sect:applications} apply the modal-trajectory uncertainty framework to assess the link between modal representations and LCSs for experimental and numerical flow field data. This includes modes obtained from: a direct numerical simulation (DNS) of the canonical cylinder flow wake at low Reynolds number \citep[$Re = 100$, see e.g.,][]{kutz_dynamic_2016,taira_modal_2020}; oscillating foil experiments \citep{Jones_Kanso_Luhar_2024}; and DNS data for a broadbanded turbulent channel flow from the Johns Hopkins Turbulence Database \citep{graham_web_2016}. 

In comparison to coherent structures obtained via FTLEs, the modal-trajectory uncertainty represents regions in the flow field where fluid particles are strongly influenced by Eulerian modes. These structures provide insight into the interactions between modes and LCSs, such as those underlying vortex shedding instabilities or large-scale turbulent structures. While the modal analysis algorithms used in this paper include POD, DMD and optimized DMD, we emphasize that this sensitivity analysis can be applied to other data-driven decompositions, including spectral POD (SPOD), balanced POD (bPOD), and operator-based methods such as resolvent analysis \citep{towne_spectral_2018,rowley_model_2017,herrmann_data-driven_2021}.

\section{Model Sensitivity}
\label{sect:msframework}
This section describes how the model sensitivity framework developed by \citet{kaszas_universal_2020} is applied to nonlinear numerical systems. The computations involved for the framework creates an upper-bound estimate for the difference between an unperturbed trajectory and a perturbed trajectory subjected to stochastic or deterministic disturbances. For the purposes of this work, we consider purely deterministic perturbations. The dynamical system for an unperturbed particle trajectory $\xb^0(t)$ is expressed as:
\begin{equation} 
\label{clean_ds} 
\dot{\xb}^0(t)=\tilde\ub(\xb^0,t),
\end{equation}
which yields the corresponding flow map:
\begin{equation}
\boldsymbol{\Phi}_{t_0}^{t}(\xb_0)=\xb_{0}+\int_{t_0}^{t}\tilde\ub(\xb^0(\tau),\tau)d\tau,\
\label{clean_t}
\end{equation}
where $\xb_{0}$ is the initial condition at time $t_0$, and $t\in[t_0, t_0 + T]$ where $T$ is the eventual period of integration. Next, we consider a deterministic perturbation $ \ub'$ to the dynamics which yields the system: 
\begin{equation}
\dot{\xb}^\epsilon(t)=\tilde\ub(\xb^\epsilon,t) + \ub'(\xb^\epsilon,t;\epsilon),
\label{pertdiff}
\end{equation}
where $\epsilon$ is a small parameter. We assume that for $0\leq\epsilon\ll1$, the perturbation depends smoothly on $\epsilon$, vanishes in the unperturbed limit and its magnitude is at most of order $\epsilon$, such that:
\begin{equation} 
|\ub'(\xb,t;0)|=0,
 \qquad
|\ub'(\xb,t;\epsilon)|\in \mathcal{O}(\epsilon)
\quad \text{as} \quad \epsilon \rightarrow 0,
\label{condition}
\end{equation}
where $|\cdot|$ represents the 2-norm. In the context of modal analysis, $\tilde\ub$ denotes a reduced-order modal representation of the velocity data, and $\ub'$ denotes the perturbation modes of interest.
\begin{figure}[t!]
    \centering    \includegraphics[width=0.8\textwidth,keepaspectratio]{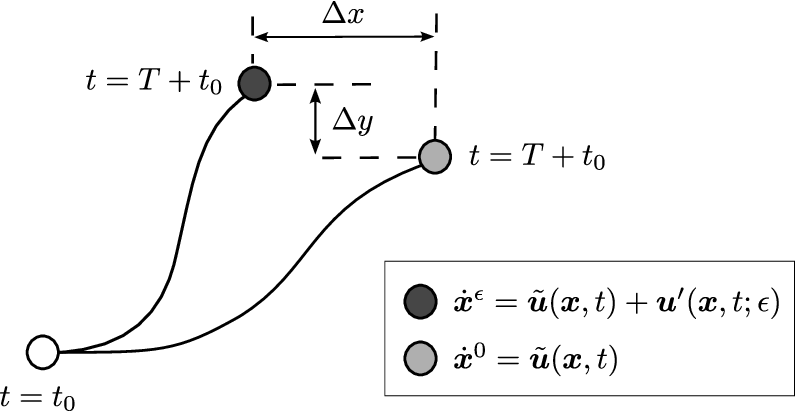}
    \caption{Illustration of perturbed (black) and unperturbed (gray) trajectories for a single fluid particle.}
    \label{fig:particle}
\end{figure}

\subsection{Definition of Upper Bound}
The leading order uncertainty of a trajectory $\xb^0(t)$ due to changes in the model can be estimated as an upper limit using quantities that relate directly to the dynamics of the system. Choosing $t = t_0+T$, as our end time, the solution can be bounded as
\begin{equation}
 \begin{split}
  \ \int_{t_0}^t |\boldsymbol{D\Phi}_s^t (\xb^0(s))||\ub'(\xb^0(s);0)| ds \leq \Delta_\infty \int_{t_0}^t \sqrt{\lambda_s^t(\xb^0(s))}ds,
 \end{split}
 \label{eq:upper_bound}
\end{equation} 
where 
\begin{equation}
\Delta_\infty=\max\limits_{s\in[t_0, \ t_0+T]}|\ub'(\xb^0(s);0)|
\label{eq:delta-inf}
\end{equation}
is the maximum value of the perturbation field ($\ub'$) along the unperturbed trajectory and $\lambda_s^t$ is the largest eigenvalue of the Cauchy-Green strain tensor $[\boldsymbol{D\Phi}_s^t]^* \boldsymbol{D\Phi}_s^t$. Here, $[\cdot]^*$ is the transpose of $[\cdot]$. 

Equation (\ref{eq:upper_bound}) is considered the universal error bound for the perturbed dynamical system shown in equation (\ref{pertdiff}) relative to the unperturbed system from equation (\ref{clean_ds}). \textbf{Model Sensitivity (MS)} with no stochastic modeling errors is defined as
\begin{equation}
\mathrm{MS}_{t_0}^t =  \left(\int_{t_0}^t \sqrt{\lambda_s^t(\xb^0(s))}ds\right)^2,
\label{eq:modelsensitivity}
\end{equation}
and is related to the mean-squared trajectory uncertainty as
\begin{equation}
|\xb^\epsilon - \xb^0|^2\leq \mathrm{MS}_{t_0}^t \Delta_\infty^2  =  \left(\int_{t_0}^t \sqrt{\lambda_s^t(\xb^0(s))}ds \Delta_\infty \right)^2.
\label{eq:sensitivity}
\end{equation}
Thus, the leading-order upper bound on the mean square trajectory uncertainty is the product of the amplitude uncertainty in the system, $\Delta_\infty$ caused by the perturbation $\ub'$, and the sensitivity with respect to initial conditions, which is characterized by the integral of the largest eigenvalue $\int_{t_0}^t\sqrt{\lambda_s^t(\xb^0(s))}ds$.

\subsection{Connection to the FTLE}
The FTLE characterizes the material line elements in a flow over a finite time interval. 
To compute FTLE fields, particle motion is integrated over a time interval $T$ using standard techniques (e.g., the Runge-Kutta method). This process yields the flow map shown in equation (\ref{clean_t}). The forward FTLE is computed as: 
\begin{equation}
\mathrm{FTLE}_{t_0}^t=\frac{1}{|T|}\log\sqrt{\lambda_{t_0}^t},
    \label{eq:fFTLE}
\end{equation}
where $\lambda_{t_0}^t$ is the maximum eigenvalue of $[\boldsymbol{D\Phi}_{t_0}^t]^* \boldsymbol{D\Phi}_{t_0}^t$. When integrated forwards in time, the FTLE field reveals unstable or repelling manifolds of trajectories. Particles can also be integrated in backward time (from $t$ to $t_0$) to reveal stable or attracting manifolds from the maximum eigenvalue $\lambda_{t}^{t_0}$. These manifolds are approximated as ridges of the backward FTLE field. 

The Model Sensitivity (MS) has a particular relationship to the forward FTLE for the baseline (unperturbed) system. Specifically, equation (\ref{eq:sensitivity}) can be rewritten as
\begin{equation}
    \frac{\log (\mathrm{MS}_{t_0}^t\Delta_\infty^2)}{2|T|} = \mathrm{FTLE}_{t_0}^t + \zeta_{t_0}^t,
    \label{eq:connection}
\end{equation}
where
\begin{equation}
    \zeta_{t_0}^t = \frac{1}{|T|}\log \left(\Delta_\infty\int_{t_0}^t{\sqrt\frac{\lambda_s^t}{\lambda_{t_0}^t}}ds \right).
    \label{eq:zeta}
\end{equation}
Thus, the upper bound on the mean-square trajectory uncertainty in equation (\ref{eq:sensitivity}), is the summation of two different effects: the sensitivity of the unperturbed system, as quantified by $\mathrm{FTLE}_{t_0}^t$, and the influence of the disturbance (via $\Delta_\infty$ in $\zeta_{t_0}^t$) on the FTLE field. Note that equation (\ref{eq:connection}) holds for the backward FTLE field if the $\mathrm{MS}$ field in equation (\ref{eq:sensitivity}) is computed in backward time as $\mathrm{MS}_t^{t_0}$.

\subsection{Example 1: Sensitivity for a model kinematic system}

We now demonstrate the $\mathrm{MS}$ framework for a simple kinematic system that is derived similarly to the model introduced by \citet{shadden_definition_2005}. An autonomous system $\dot{\xb}=\tilde\ub(\xb)$ is constructed with
\begin{equation}
    \tilde\ub(\xb)= \left[ \begin{matrix}
         U_0-\pi A \sin(\pi x)\cos(\pi y) \\  \pi A \cos(\pi x)\sin(\pi y) 
         \end{matrix} \right].
\end{equation}
In the above equations, $A$ is the amplitude of the spatially-varying velocity field, while $U_0$ represents a background flow in the model, so that the particles primarily move in the positive $x$ direction. The velocity parameters are set to $A=0.1$ and $U_0=0.2$. These baseline parameters were chosen based on prior literature. Notably, the resulting flow comprises two counter-rotating gyres and yields manifold structures similar to those found in \citep{rom-kedar_analytical_1990} which are relevant to a range of applications, including wavy-walled channel flows and trailing edge vortices. Following equation (\ref{pertdiff}), we introduce a crossflow perturbation $\ub'(\xb,t;\epsilon)$ in the form of a jet that oscillates in the $y$ direction:
\begin{equation}
    \ub'(\xb,t;\epsilon)=\left[ \begin{matrix} 0 \\  \epsilon\sech^2(c(x-d)) \sin(\omega t)
    \end{matrix} \right],
\end{equation}
where $c$ and $d$ are parameters describing the jet width and position respectively. This perturbation can be viewed as a dynamical mode superimposed on the base system, selectively exciting transverse motions on top of $\tilde\ub(\xb)$. The non-autonomous dynamical system $\dot{\xb}=\tilde\ub(\xb)+\ub'(\xb,t;\epsilon)$ also satisfies the continuity equation. We choose the spatial parameters $c=4$ and $d=1.4$, and a frequency of $\omega=2\pi/5$, i.e., the dimensionless oscillation period is $T_p=5$. For the FTLE and $\mathrm{MS}$ calculations presented below, the grid and time resolutions were $\Delta x=0.006$, $\Delta t=0.025$ and the integration period was set to $|t-t_0|=T=15$, representing three oscillation periods for the perturbation field.

\begin{figure}[t!]
    \centering    \includegraphics[width=0.52\textwidth,keepaspectratio]{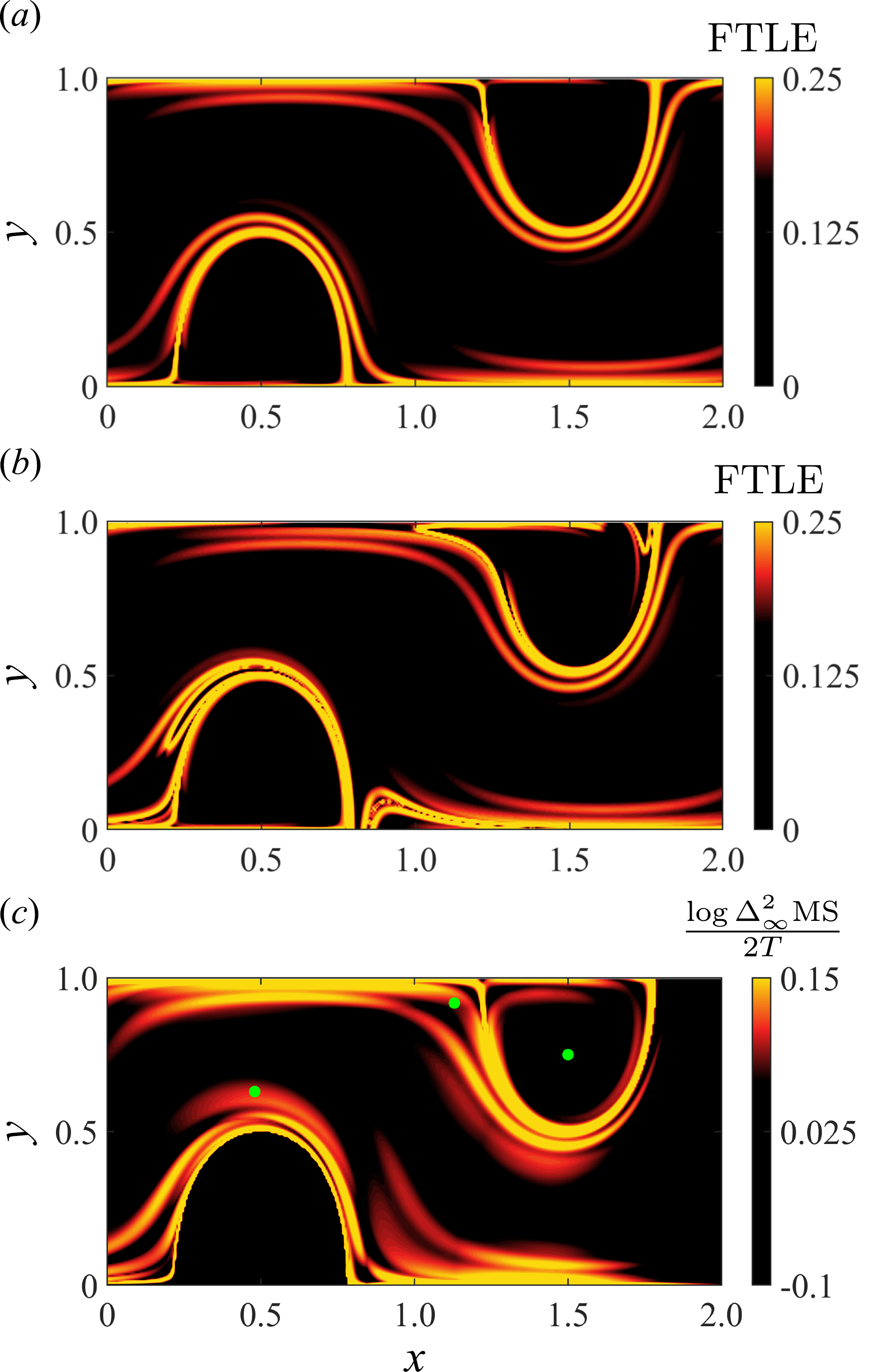}
    \caption{($a$) Forward FTLE field showing LCS structure of the unperturbed kinematic system with $\epsilon=0$. ($b$) LCS structure of the perturbed kinematic system $\tilde\ub(\xb)+\ub'(\xb,t;\epsilon)$ for inital time $t_0=T/2$ with $\epsilon=0.03$. ($c$) Scaled upper-bound uncertainty fields for the perturbed kinematic system for initial time $t_0=T/2$ with $\epsilon=0.03$. The green points are particle positions referred to in figure \ref{fig:currentbounds}.}
    \label{fig:currentall}
\end{figure}

Figure \ref{fig:currentall}$a$ shows the forward FTLE field for the unperturbed system $\tilde\ub(\xb)$ on the non-dimensional grid $[0,2] \times [0,1]$. Two dominant  horseshoe ridges centered at $x=0.5$ (lower left) and $x=1.5$ (upper right) are observed. Fluid particles outside the horseshoe boundaries generally travel in the positive $x$ direction due the background flow $U_0$. In contrast, particles inside the horseshoe structures do not exit and continuously circulate. 

Applying the small perturbation $\ub'(\xb,t;\epsilon)$ with $\epsilon=0.03$, as in figure \ref{fig:currentall}$b$, moderately alters the repelling FTLE manifolds, producing additional loops around the lower-left horseshoe and deformations of the upper-right horseshoe, while still retaining the dominant FTLE field features observed in figure \ref{fig:currentall}$a$. Regions where the LCS is influenced by the perturbation are indicated by the upper-bound uncertainty field in figure \ref{fig:currentall}$c$. High values of $\Delta^2_{\infty}\mathrm{MS}$ coincide with the lobe perturbations observed for the lower-left horseshoe, whereas large values near the upper-right boundary indicate the deformation of the manifolds. 

\begin{figure}[t!]
    \centering    \includegraphics[width=0.97\textwidth,keepaspectratio]{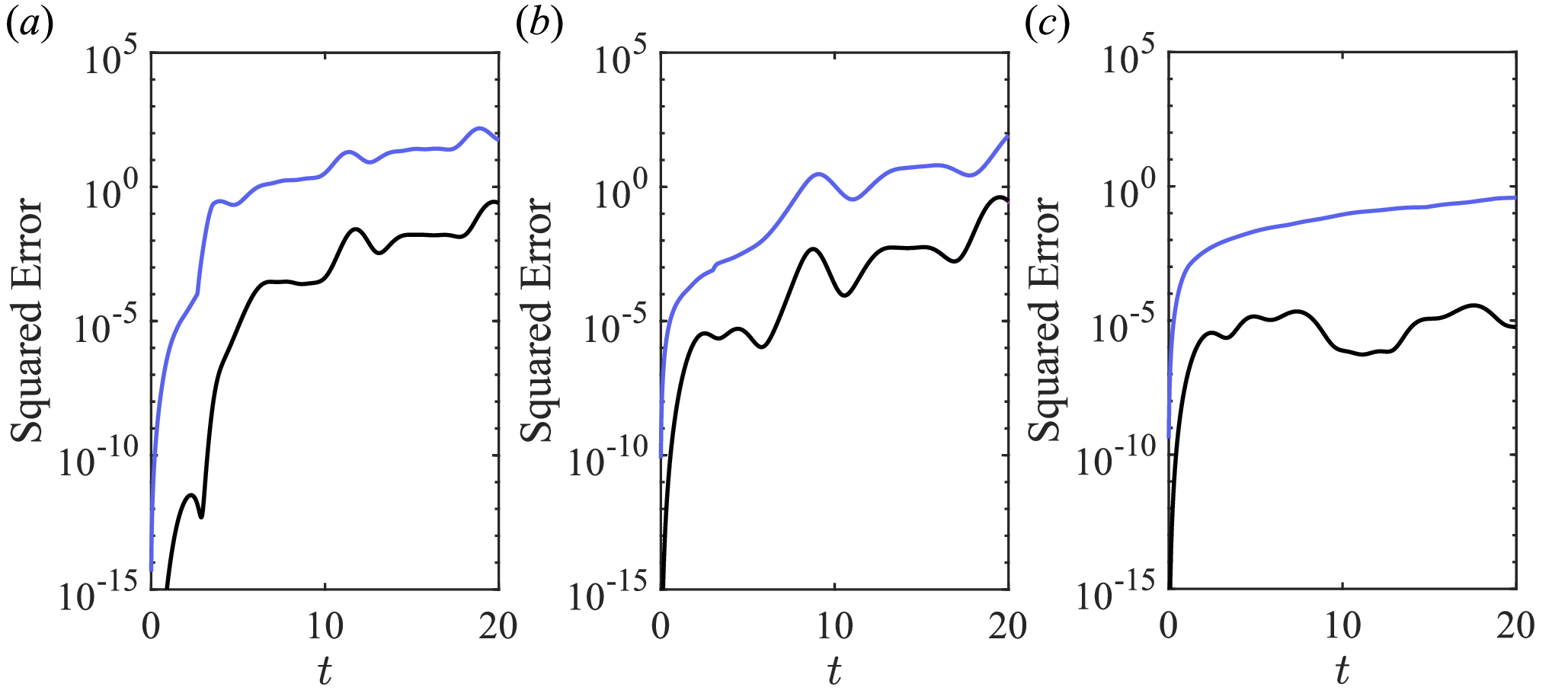}
    \caption{ Trajectory uncertainty (black) described by $|\xb^\epsilon - \xb^0|^2$ and upper-bound estimate (blue) of trajectory uncertainty represented by $\Delta^2_\infty \mathrm{MS}$. The initial conditions ($x,y$) chosen for each of the curves are ($a$) ($0.48, 0.63$), ($b$) ($1.13, 0.92$) and ($c$) ($1.50, 0.75$).  These positions are marked with green points in figure~\ref{fig:currentall}$b$.}
    \label{fig:currentbounds}
\end{figure}

Particle trajectory errors are compared to their upper-bound estimates in figure \ref{fig:currentbounds}. The upper bound estimates capture trends in the trajectory uncertainty well, though the true trajectory deviations are two to four orders of magnitude lower. The upper estimates tightly follow the dynamics of the trajectory errors, particularly at initial positions with high values of $\Delta^2_\infty \mathrm{MS}$, as shown in figures \ref{fig:currentbounds}($a,b$). In contrast, particles inside the horseshoe structures, such as those in the region shown in figure \ref{fig:currentbounds}($c$), undergo weaker relative stretching and therefore do not track the uncertainty errors as closely. Broadly, the dynamics in the trajectory uncertainty can be predicted well when the upper-bound estimates are computed in locations near or directly on the FTLE ridges.


\section{Modal-trajectory uncertainty}
\label{sect:modal-trajectory uncertainty}
We now consider a particular definition of model sensitivity in the context of modal representations derived from data. Consider a set of snapshots from a velocity field, $\ub(\xb,t)$, which can have either two or three-dimensions. The general aim of modal representations is to take the high dimensional dataset $\ub$ and compress the data into a lower dimensional form $\hat{\ub}$ that represents the main dynamics. Thus, $\hat{\ub}$ is expressed as a limited set of $r$ linearly independent modes (i.e., a rank $r$ approximation):
\begin{equation}
    \ub \approx \hat{\ub }= \sum_{k\in\mathcal{H}} a_k \boldsymbol{\psi}_k,
    \label{eq:linmode}
\end{equation}
where $\boldsymbol{\psi}_k$ is a basis function with corresponding amplitude coefficient $a_k$, and $\mathcal{H}$ denotes the set of all modes ($\mathcal{H}=\{0,1,\ldots,r\}$). Variants of the decomposition methods described by equation (\ref{eq:linmode}) lead to different basis functions $\boldsymbol{\psi}_k$. Proper orthogonal decomposition (POD) \citep{berkooz_proper_1993}, for example, yields orthonormal basis functions for the modes, which follow the condition:
\begin{equation}
    \langle \boldsymbol{\psi}_i, \boldsymbol{\psi}_j \rangle = \int_{\Omega} \boldsymbol{\psi}_i(\xb) \boldsymbol{\psi}_j(\xb) \, d\xb = 
\begin{cases} 
1 & \text{if } i = j \\ 
0 & \text{if } i \neq j 
\end{cases},
\end{equation}
where $\Omega$ is the full spatial domain of the dataset. Alternatively, dynamic mode decomposition (DMD) as introduced by \citet{schmid_dynamic_2010} yields basis functions that contain a particular frequency and growth rate. For a reconstruction of $\ub$ based on DMD modes, equation (\ref{eq:linmode}) becomes:
\begin{equation}
     \hat{\ub}=\sum_{k\in\mathcal{H}} b_k \exp(\omega_k t)\boldsymbol{\psi}_k,
    \label{eq:dmdmode}
\end{equation}
where $b_k$ is the mode amplitude coefficient and $\omega_k$ is the eigenvalue corresponding to the eigenvector $\boldsymbol{\psi}_k$.  A more detailed overview of modal analysis techniques is provided by \citet{taira_modal_2020} and \citet{rowley_model_2017}.

The modal approximation in equation \ref{eq:linmode} can be partitioned into two linear combinations,
$\tilde{\ub}(\xb,t)$ and $\ub'(\xb,t)$, corresponding to two disjoint
subsets of modes, $\mathcal{F}$ and $\mathcal{G}$, respectively:
\begin{equation}
    \bm{u}(\bm{x},t)
\approx
\sum_{i \in \mathcal{F}}
a_i(t)\,\boldsymbol{\psi}_i(\bm{x})
+
\sum_{j \in \mathcal{G}}
a_j(t)\,\boldsymbol{\psi}_j(\bm{x})=\tilde{\ub}(\xb,t)+\ub'(\xb,t),
\label{eq:CSeqn}
\end{equation}
where
$\mathcal{G}\cup\mathcal{F}=\mathcal{H}$, and
$\mathcal{G}\cap\mathcal{F}=\emptyset$.
Note that equation~(\ref{eq:CSeqn}) has a similar form to the parameterized system in equation~(\ref{pertdiff}). However $\ub'(\xb,t)$ in the above equation represents a small perturbation constructed from the modes in $\mathcal{G}$, with its magnitude contained in the corresponding modal coefficients $a_j(t)$ rather than by an explicit parameter $\epsilon$. Another key distinction is that the mode components in equation~(\ref{eq:CSeqn}) can be defined in various ways, provided that $\ub'$ remains sufficiently small so that $\tilde{\ub}$ captures the dominant dynamics of the full flow field $\ub$ $\left ( |\ub'|/|\tilde{\ub}| \ll 1 \right)$. For example, $\ub'$ can be chosen to represent the modes neglected in a reduced-order model or a subset of modes that relate to specific flow features.

 Since the dynamics of passive tracer particles in the flow field are given by:
\begin{equation}
    \frac{d\xb}{dt}=\ub(\xb,t),
\end{equation}
it follows that the model sensitivity framework developed by \citet{kaszas_universal_2020} can be used to characterize how the Eulerian modes in $\ub'(\xb,t)$ influence the LCSs estimated using the baseline system $\tilde{\ub}$. For simplicity, we use the term \textbf{modal-trajectory uncertainty} ($\Delta^2_\infty \mathrm{MS}$) to refer to the leading-order trajectory uncertainty estimate of equation (\ref{eq:sensitivity}). This quantity represents the combined product of the model sensitivity (MS) and the mode perturbation fields ($\Delta_\infty$). Furthermore, the term $\zeta$ in equation (\ref{eq:connection}) can be thought of as a modal perturbation to the FTLE field. 


For the unsteady shear flows considered in this paper, most decomposition methods explicitly yield a mean (or time-averaged) flow as one of the leading modes. Often, the mean flow contains more energy (in the $L_2$ sense) than other modes. Thus, for all cases in this study, we will consider the mean flow $\bar{\ub}$ as part of the modal representation $\tilde{\ub}(\xb,t)$.

\subsection{Computing the modal-trajectory uncertainty field}
\label{sect:compute-modal-uncertainty}
Often, flow fields acquired from numerical simulations or experiments have a limited interrogation region. Pertinent to the FTLE and $\Delta_\infty ^2 \mathrm{MS}$  calculations, this affects how long passive tracers are advected before exiting the domain. One approach for circumventing this issue is to select a subdomain to initialize the passive tracers in the existing data. However, this is typically not a full solution since the integration $T$ must be long enough for a sufficient number of particles to exit the subdomain. Another practical approach is to assign particles that exit the domain an approximation of the velocity, based on the appropriate boundary conditions from the data \citep{rockwood_practical_2019}.

For flow fields that have a dominant advection direction, passive tracers that exit a distance $\Delta \xb$ from the domain can continue to advect based on the mean flow field, expressed as:
\begin{equation}
    \ub(\xb+\Delta \xb,t) = \bar{\ub}(\xb,t).
    \label{eq:mean}
\end{equation}
Alternatively, one may use Taylor's frozen eddy hypothesis \citep{taylor_spectrum_1997}, which assumes that a passive tracer that exits the domain at a distance $\xb+\Delta \xb$ and time $t$ experiences the same velocity at a distance $\xb$ and earlier time $t-\frac{\Delta x}{u_a}$; that is:
\begin{equation}
    \ub(\xb+\Delta \xb,t) \approx \ub\left(\xb,\ t-\frac{\Delta x}{u_a(\xb,\ t)}\right),
    \label{eq:taylor}
\end{equation}
where $u_a$ is an advection speed that is chosen based on the characteristic flow velocity, such as the local mean velocity or the peak velocity within the domain. This approximation effectively treats turbulent structures as "frozen" as they are advected past a point, allowing temporal measurements at a fixed location to be interpreted as spatial information along the direction of the flow. In the following sections, we compute the Lyapunov exponent $\lambda_s^t$ in equation (\ref{eq:sensitivity}) using equation (\ref{eq:mean}) to approximate particle trajectories that exit the domain. Similarly, the computation of $\Delta_\infty$ also requires a flow field approximation for particles that leave the domain due to the limited spatial data of the perturbation $\ub'(\xb,t)$. Unless stated otherwise, the values of $\ub'(\xb,t)$ were extrapolated using Taylor's frozen eddy hypothesis as in equation (\ref{eq:taylor}), where $u_a(\xb,\ t)$ is the streamwise velocity at the particular lateral location.

\section{Applications and Results} \label{sect:applications}
\subsection{Example 1: Flow past the wake of a circular cylinder}

\begin{figure}[t!]
    \centering  \includegraphics[width=0.75\textwidth,keepaspectratio]{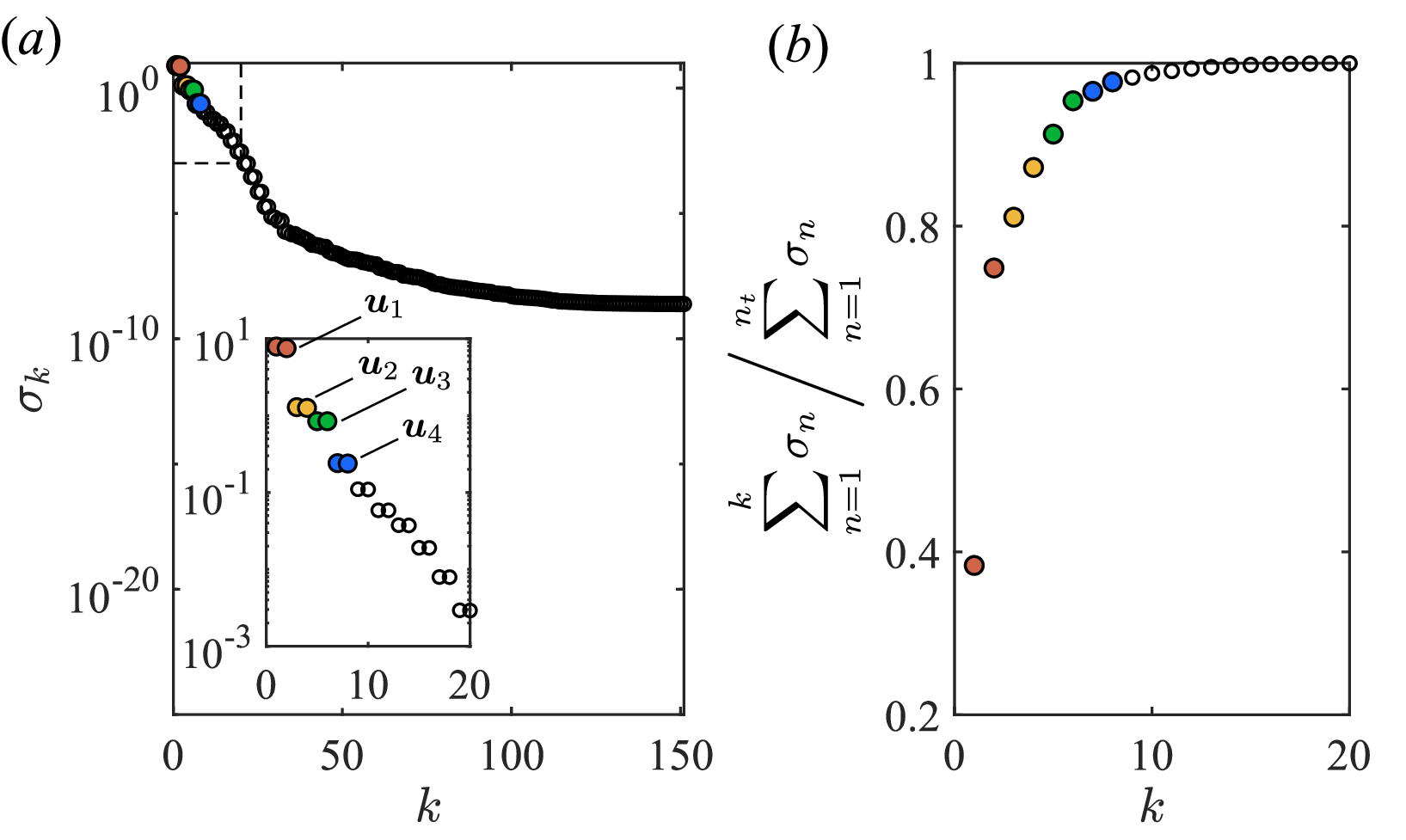}
     \caption{($a$) Singular values corresponding to modes obtained via snapshot POD and ($b$) cumulative energy spectrum for the cylinder wake. The inset in ($a$) highlights the mode pairs in red ($\bm{u}_1$), orange ($\bm{u}_2$), green ($\bm{u}_3$), and blue ($\bm{u}_4$).}
    \label{fig:cyl-spect}
\end{figure}

We first show how the modal-trajectory uncertainty field is useful for the analysis of vortex-dominated flows. In particular, we use the numerical data of the canonical cylinder wake flow ($Re=u_\infty d/\nu=100$ and $St=0.16$) from \citet{Boudina2021}, where $d$ is the cylinder diameter. Numerical results were obtained using Cadyf, a fluid-structure interaction solver based on the finite element method \citep{etienne_perspective_2009}. The original velocity data were defined on an unstructured mesh and interpolated onto a uniform Cartesian grid with dimensionless resolutions of $\Delta x=0.08$ and $\Delta y=0.06$. The interpolated velocity fields were obtained over the subdomain $(x,y) \in [-21,\ 20] \times [-8,\ 8]$ and constituted a grid of $525 \times 266$ velocity vectors. The mean flow ($\bar{\bm{u}}$) was subtracted from the velocity data before applying POD for the modal-trajectory uncertainty computations. 

\begin{figure}[t!]
    \centering  \includegraphics[width=0.70\textwidth,keepaspectratio]{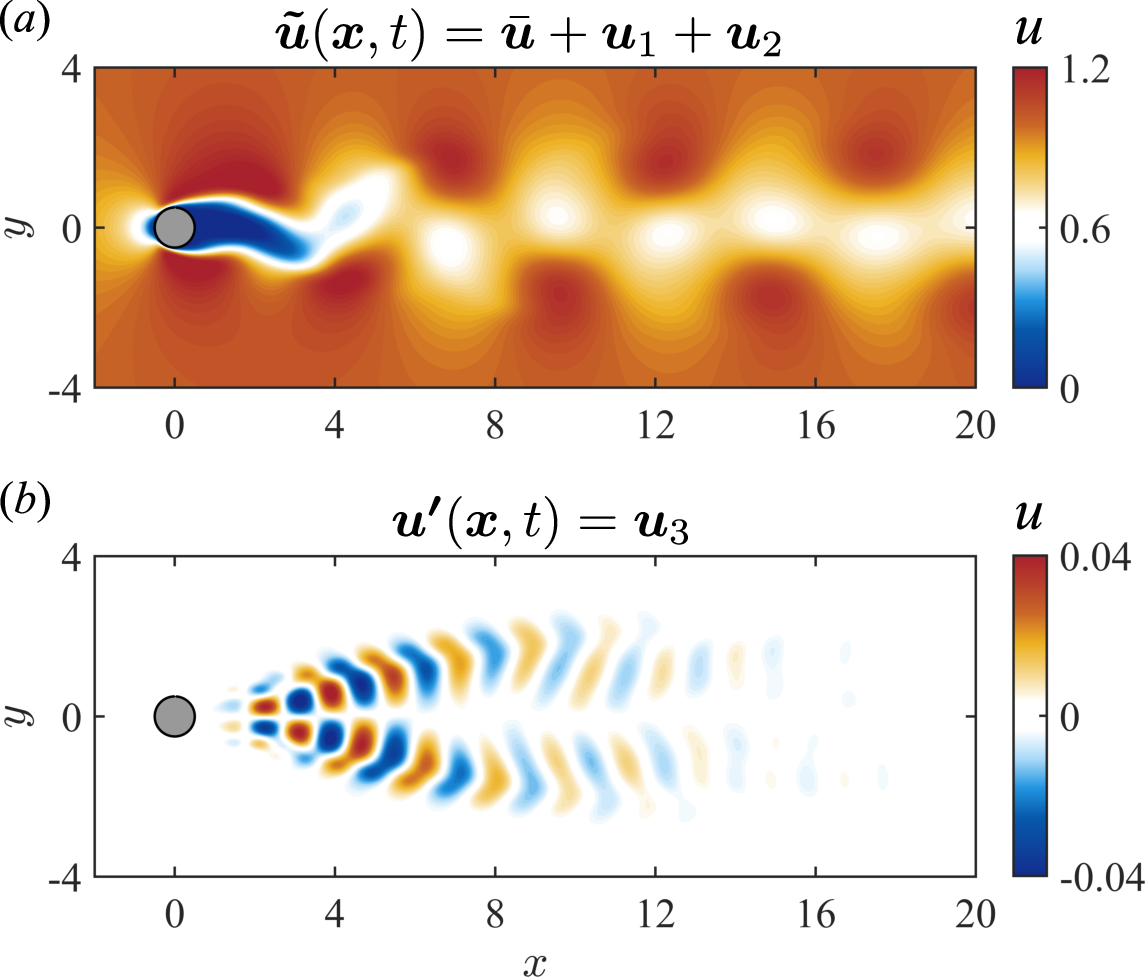}
    \caption{Streamwise velocity of the modal representation $\tilde{\ub}(\xb,t)$ ($a$) and perturbations $\ub'(\xb,t)$ ($b$) used for the modal-trajectory uncertainty of the cylinder wake.}
    \label{fig:cyl-dyn}
\end{figure}

The spectrum of POD modes is shown in figure \ref{fig:cyl-spect}. The mode pair with the highest energy (denoted as $\ub_1$) corresponds to the shedding frequency of the wake, while subsequent modes ($\ub_2$, $\ub_3$ etc.) are higher-order harmonics \citep{noack_hierarchy_2003}. To better understand how the harmonic modes affect the vortex street, we consider the following modal representation and perturbation for the modal-trajectory uncertainty analysis:
\begin{equation}
    \tilde{\ub}(\xb,t) = \bar{\ub}(\xb)+\ub_1(\xb,t) + \ub_2(\xb,t), \ \ \ \ub'(\xb,t) = \ub_3(\xb,t),
    \label{cyl-sys2}
\end{equation}
which are shown in figure \ref{fig:cyl-dyn}. The relative magnitude of the perturbation was estimated as $|\ub'|/|\tilde{\ub}| \approx 0.0034$ using an $L_2$ norm, indicating that the modes contained in $\ub'$ provide a small perturbation to the baseline flow $\tilde{\ub}$. The FTLE and modal-trajectory uncertainty fields were computed in backward time over the subdomain $(x,y) \in [-2,\ 20] \times [-5,\ 5]$. Tracer particles that exited this subdomain through the $x$-boundaries were advected using the mean-field approximation in equation~\ref{eq:mean}. No tracer particles exited the subdomain through the
$y$-boundaries.


\begin{figure}[t!]
    \centering  
     \includegraphics[width=0.70\textwidth,keepaspectratio]{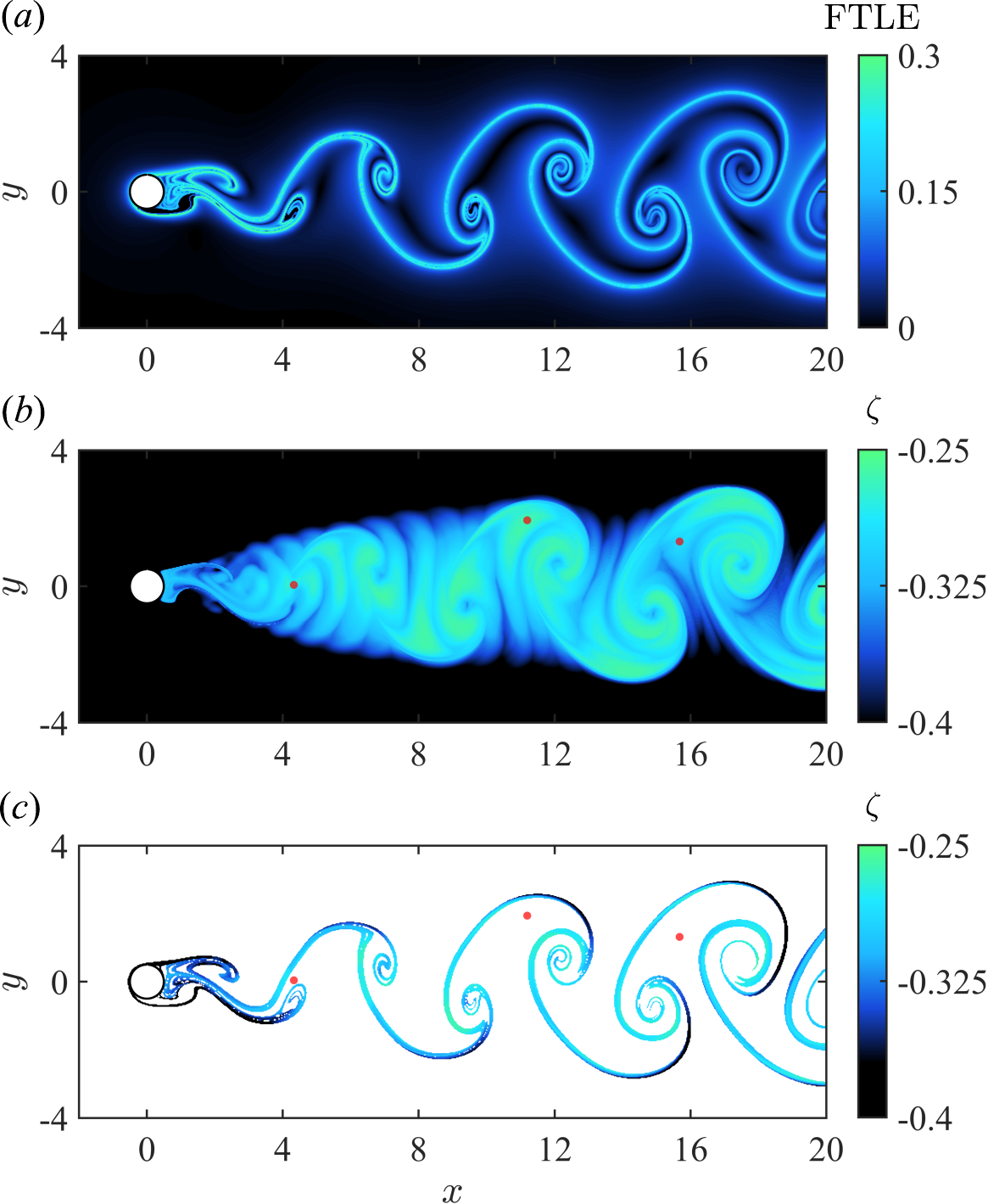}
    \caption{ Lagrangian coherent structures of the cylinder wake. ($a$) Backward FTLE structure original flow field for the cylinder wake. ($b$) FTLE perturbation $\zeta$ for the cylinder wake flow field. ($c$) Magnitudes of $\zeta$ superposed onto the backward FTLE field for the baseline flow field $\tilde{\ub}(x,t)$. Backward FTLE values that were less than 38\% of the maximum were removed. Each row corresponds to the flow structure cases in figure \ref{fig:cyl-dyn}. The red points are the regions referred to in figure \ref{fig:cyl-bounds}.}     
    \label{fig:cyl-zeta}
\end{figure}

The backward FTLE field of the original flow field is shown in figure \ref{fig:cyl-zeta}($a$), illustrating the dynamics of the von Karman vortex street. Figure~\ref{fig:cyl-zeta}($b$) shows the corresponding FTLE perturbation, $\zeta$, which combines spatial features identified by the backward FTLE field and the Eulerian mode perturbations in figure~\ref{fig:cyl-dyn}($b$). The FTLE perturbation further illustrates how modal perturbations advect downstream and interact with the vortex street, with the strongest alterations to the vortex roll-up occurring in the far wake region $x\in[8,20]$. To identity where these perturbations overlap with the dominant manifolds, we superimpose the response $\zeta$ onto the FTLE field for the baseline system $\tilde{\ub}(x,t)$, as illustrated in figure~\ref{fig:cyl-zeta}($c$). The highest values of $\zeta$ are concentrated towards the vortex lobes of the FTLE manifolds, indicating that particle trajectories near these regions are most strongly influenced by the harmonic mode perturbations. This observation is consistent with the modes being the most energetic in the near-wake ($1\leq x < 8$ in figure \ref{fig:cyl-dyn}$b$) while their influence on Lagrangian coherent structures extends several diameters downstream  ($8\leq x \leq 20$ in figure \ref{fig:cyl-zeta}$c$).
\begin{figure}[t!]
    \centering  \includegraphics[width=0.55\textwidth,keepaspectratio]{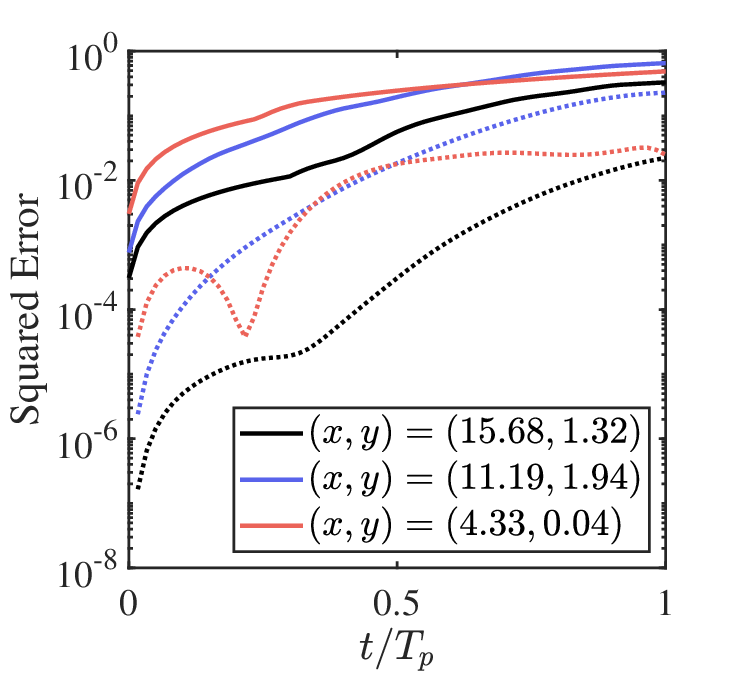}
    \caption{Upper bound estimates $\Delta_\infty^2 \mathrm{MS}_{t_0}^t$ (bold) and trajectory uncertainties $|\xb^\epsilon - \xb^0|^2$ (dotted) for the cylinder wake flow. The initial values of $\zeta(\bm{x}_0)$ at the positions denoted by the red markers in figure \ref{fig:cyl-zeta}$b$ for each of the curves are: black ($\zeta = -0.30$);
    blue ($\zeta =-0.26) $; and 
    red ($\zeta = -0.28$).} 
    \label{fig:cyl-bounds}
\end{figure}
\\
A correlation is also observed between the FTLE perturbation values and the modal-trajectory uncertainty, as shown in figure \ref{fig:cyl-bounds}. Initial points with large perturbation values $\zeta$ tend to produce higher trajectory errors than those with smaller $\zeta$. Additionally, within the first oscillation period, and particularly for $t/T_p<0.5$, the $\Delta_\infty^2\mathrm{MS}$ curves closely track the growth of the trajectory error, consistent with the behavior observed for the kinematic model in figure \ref{fig:currentbounds}. Although exploratory, these findings highlight that the FTLE perturbation field $\zeta$ and modal-trajectory uncertainty can be used to predict how strongly fluid particles are influenced by specific mode structures.
\subsection{Example 2: Flow past the wake of an oscillating foil}
\begin{figure}[t!]
    \centering  \includegraphics[width=0.55\textwidth,keepaspectratio]{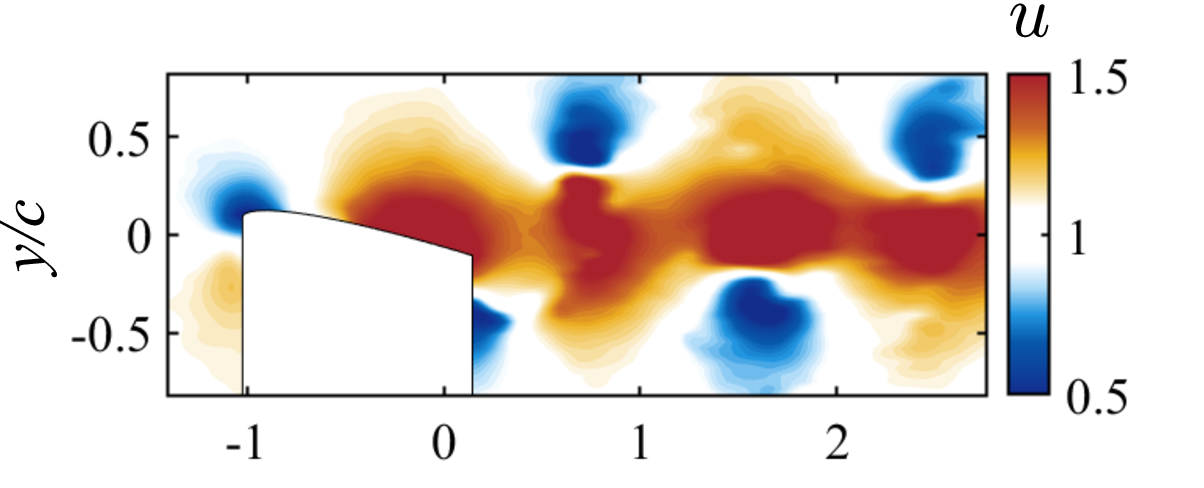}
    \caption{($a$) Streamwise velocity field from an oscillating foil ($St=0.29$) at the phase $t=T_p/4$ }
    \label{fig:osc-orig-u}
\end{figure}

\begin{figure}[t!]
    \centering  \includegraphics[width=0.75\textwidth,keepaspectratio]{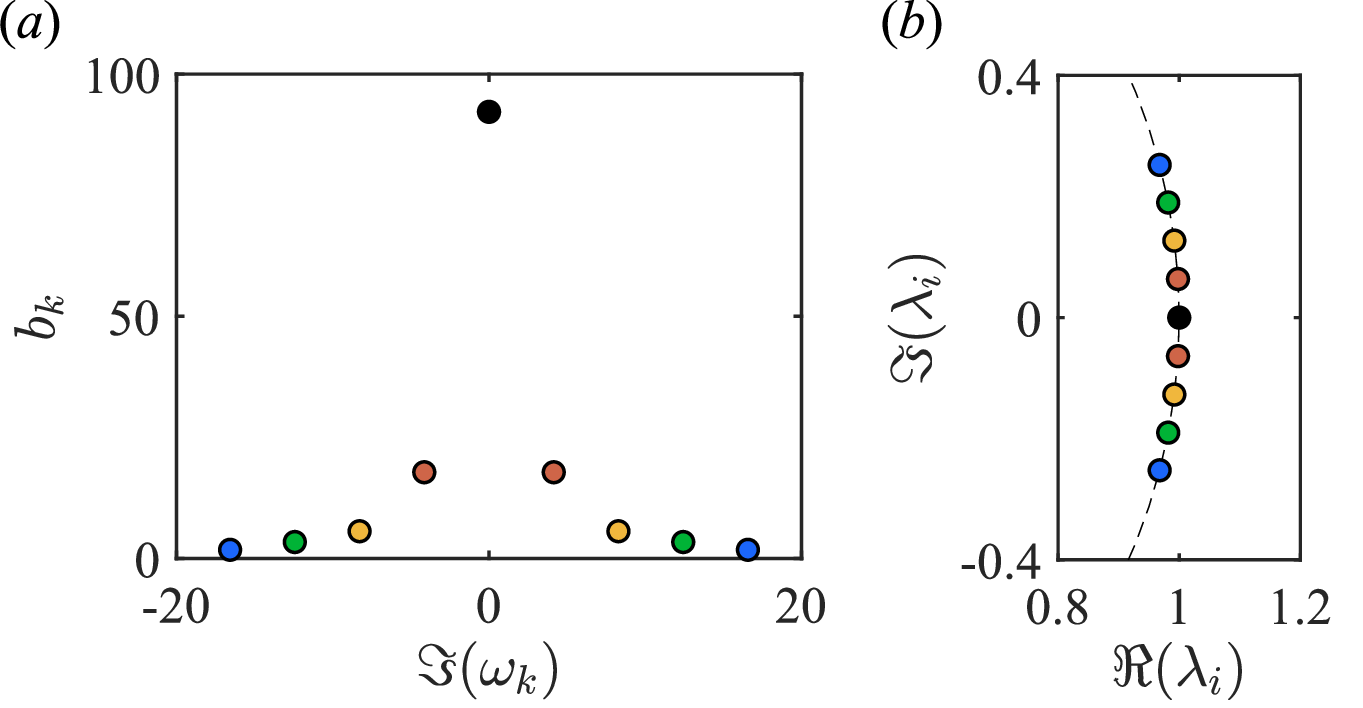}
    \caption{ ($a$) Amplitude and ($b$) eigenvalue spectrum of the opt-DMD modes for the flow past an oscillating foil. The leading unsteady mode pairs are highlighted in red ($\ub_1$), orange ($\ub_2$), green ($\ub_3$), and blue ($\ub_4$). The mean flow highlighted in black is denoted as $\bar{\ub}$.}
    \label{fig:osc-spect}
\end{figure}

We now evaluate modal-trajectory uncertainty for particle image velocimetry (PIV) data in the wake of a NACA-0012 oscillating foil at a chord Reynolds number $Re_c=U_{\infty}c/\nu =11\times10^3$ and a Strouhal number of $St=\omega A/2\pi U_\infty=0.29$ from \citet{Jones_Kanso_Luhar_2024}. Here, $\omega$ is the oscillation frequency, $A$ is the amplitude of oscillation at the trailing edge, $c$ is the foil chord length, $U_\infty$ is the freestream velocity and $\nu$ is the kinematic viscosity. For fixed heave and pitch amplitudes, and a constant phase difference between the sinusoidal pitching and heaving motions, this oscillation frequency results in peak propulsive efficiency. PIV data were obtained in the wake of the oscillating foil over the domain $(x/c, \ y/c) \in [0.35, \ 2.76] \times [-0.82,\ 0.82]$, where the coordinates are relative to the tip of the foil at pitch angle and heave position of $0^\circ$ and $0$ respectively. An additional field-of-view centered on the foil was used to better capture the wake structures produced around the foil. A dynamic mask was used to remove erroneous velocity vectors within or near the foil region. After removing erroneous velocity vectors, both datasets were combined in post-processing using a moving-average filter to reduce the signal-to-noise ratio. The combined data extended the PIV domain to $(x/c, \ y/c) \in [-2, \ 2.76] \times [-0.82,\ 0.82]$ and contained $141 \times 49$ velocity vectors. The streamwise component of the combined flow field is shown in figure \ref{fig:osc-orig-u}.

\begin{figure}[t!]
    \centering  \includegraphics[width=1\textwidth,keepaspectratio]{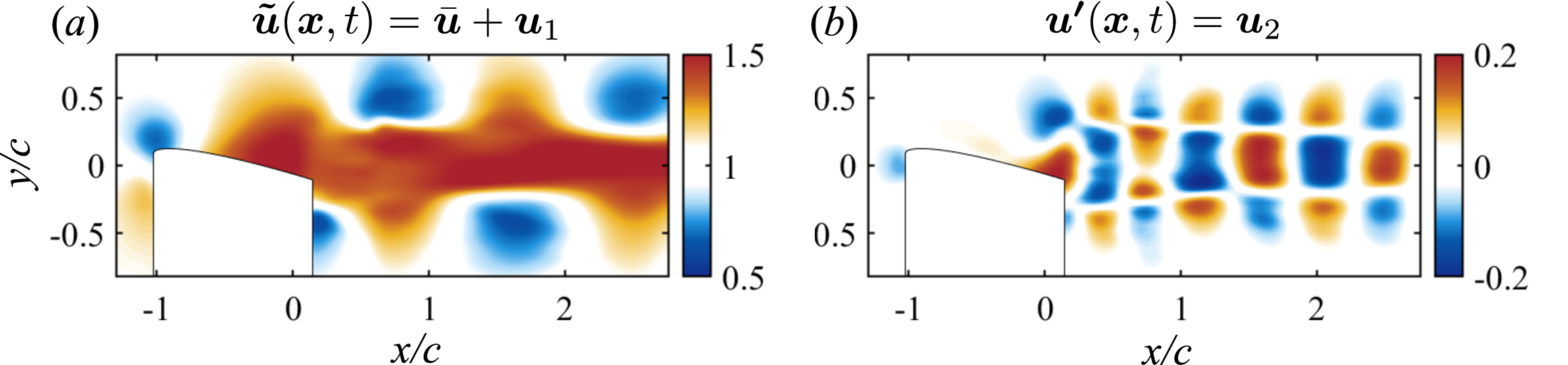}
    \caption{Streamwise component for (a) the modal representation and (b) mode perturbation for the flow past an oscillating foil at the phase $t=T_p/4$.}
    \label{fig:osc-dyn}
\end{figure}

\begin{figure}[t!]
    \centering  \includegraphics[width=0.55\textwidth,keepaspectratio]{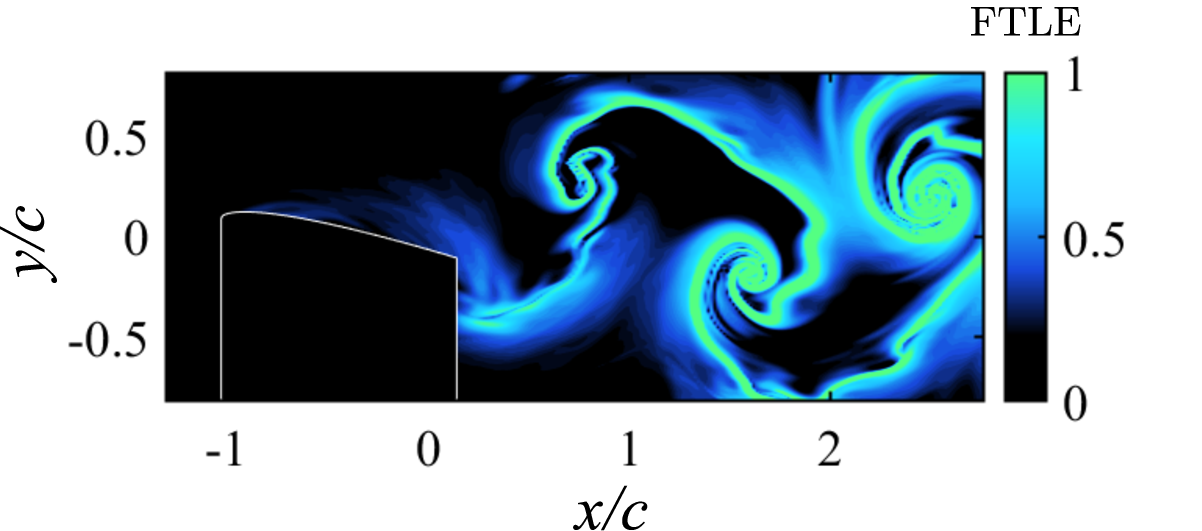}
    \caption{Backward FTLE structure of the original flow field described in figure \ref{fig:osc-orig-u} at the phase $t=T_p/4$.}
    \label{fig:osc-orig-ftle}
\end{figure}

\begin{figure}[t!]
    \centering  \includegraphics[width=1\textwidth,keepaspectratio]{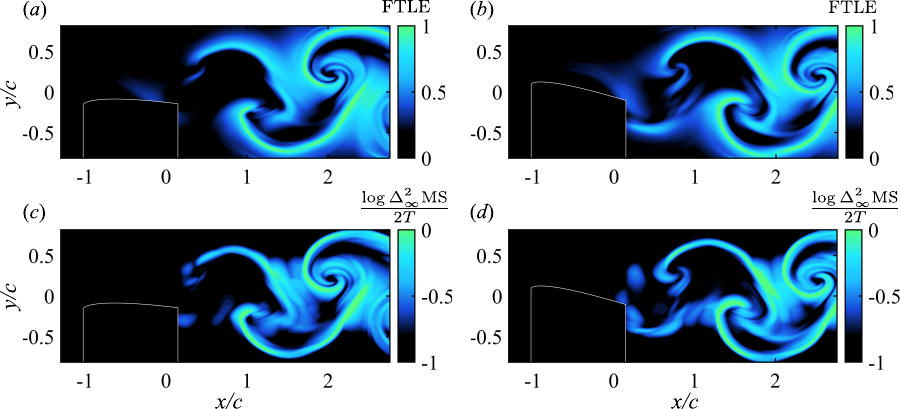}
    \caption{LCSs of the flow past an oscillating foil for the system $\tilde{\ub}(\xb,t)$ ($a,b$) as described in equation (\ref{osc-sys1}) and ($c,d$) modal-trajectory uncertainty fields at the phases (left) $t=0$ and (right) $t=T_p/4$.}
    \label{fig:osc-ftle-pert-a}
\end{figure}

\begin{figure}[t!]
    \centering  \includegraphics[width=1\textwidth,keepaspectratio]{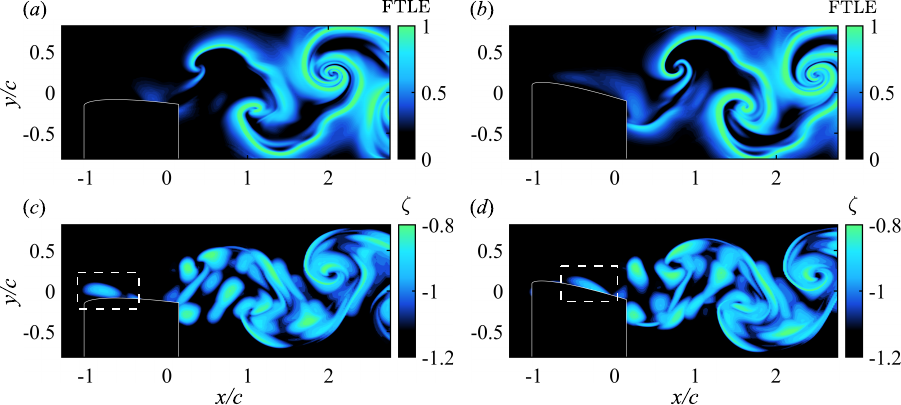}
    \caption{LCSs of the flow past an oscillating foil for the system $\tilde{\ub}(\xb,t)+\ub'(\xb,t)$ ($a,b$) as described in equation (\ref{osc-sys1}) and ($c,d$) FTLE perturbation fields at the phases (left) $t=0$ and (right) $t=T_p/4$. The dashed box highlights a component of the leading edge vortex.}
    \label{fig:osc-ftle-pert-b}
\end{figure}

The aim in this example is to understand how particular modes that relate to high propulsive efficiency affect the LCS structure. In particular, previous work \citep{Jones_Kanso_Luhar_2024} showed that optimized-DMD (opt-DMD) modes obtained from this flow field highlight thrust and drag producing effects at different Strouhal numbers, $St$. Here, we apply the same opt-DMD method to the combined dataset to extract the corresponding mode structures and examine how they influence the FTLE wake structure through the modal-trajectory uncertainty framework. The opt-DMD method, as described by \citep{askham_variable_2018}, fits the reconstruction in equation (\ref{eq:dmdmode}) by solving a nonlinear least-squares problem for the eigenvalues and respective amplitude coefficients. The spectrum of opt-DMD modes is shown in figure \ref{fig:osc-spect}. The four leading unsteady modes, ordered in terms of frequency, are denoted $\ub_1 - \ub_4$. Consistent with the previous study, the primary mode $\ub_1$ for this dataset corresponds to the oscillation frequency of the foil and, together with the mean flow $\bar{\ub}$ (figure \ref{fig:osc-dyn}$a$), represents a majority of the wake dynamics. The secondary mode $\ub_2$, as shown in figure \ref{fig:osc-dyn}$b$, contributes to the shear layer roll up of the vortices. 


For this Strouhal number case, both $\ub_1$ and $\ub_2$ yield positive induced velocities in the wake, contributing to thrust generation and high efficiency. Because we are interested in characterizing the effect of mode $\ub_2$ on the dominant LCS structure of the wake, we choose the unperturbed dynamical system and the perturbation to be
\begin{equation}
    \tilde{\ub}(\xb,t) = \boldsymbol{\bar{\ub}}(\boldsymbol{x}) + \ub_1(\xb,t), \ \ \  \ub'(\xb,t) = \ub_2(\xb,t),
    \label{osc-sys1}
\end{equation}
as illustrated in figure \ref{fig:osc-dyn}. The relative magnitudes between the modal representations were $|\ub'|/|\tilde{\ub}| \approx 0.024$. The LCS and modal-trajectory uncertainty fields for this dataset are computed in backward time as $\mathrm{FTLE}_{t}^{t_0}$ and $\mathrm{MS}_{t}^{t_0}$ as in the cylinder wake example. Given the uniform flow field upstream, we assume that particles that exit the limited domain in the negative $x-$direction and either $y-$direction experience a perturbation of $\ub'=0$, and a normalized freestream velocity of $u_\infty=1$. These assumptions are used to approximate the amplitude $\Delta_\infty$.

The backward FTLE field for the original (full) flow field is shown in figure \ref{fig:osc-orig-ftle}. Vortex roll-up is observed where the manifolds spiral inwards, as also found from the results of \citet{green_unsteady_2011}. The FTLE field for the baseline (unperturbed) velocity field is shown in figures \ref{fig:osc-ftle-pert-a}$(a,b)$, which comprises the mean flow and the first leading modal contribution. Weakened vortex rollers are observed in this reduced model. As expected, the modal-trajectory uncertainty fields in figures \ref{fig:osc-ftle-pert-a}$(c,d)$ effectively reflect the dynamics of original FTLE field. Specifically, the $\Delta_\infty ^2\mathrm{MS}$ field highlights the lower spiral which is weakened in the reduced model as well as the sections of the shear layers that show reduced undulation. These effects from the secondary mode $\ub_2$ on the wake structure are further highlighted in figures \ref{fig:osc-ftle-pert-b}$(a,b)$, where adding the mode strengthens the roll-up and leads to increased corrugation of the shear layer, consistent with features observed in the full FTLE field. 
 
Similarly to the observations for the cylinder wake case, the response field $\zeta$ (figures \ref{fig:osc-ftle-pert-b}$c,d$) provides additional insight into how mode $\ub_2$ relates to the flow structure. In particular, components of the leading-edge vortex are observed near the foil and advect downstream into the LCS structure. This feature is not immediately evident in any of the FTLE fields presented. The FTLE perturbation $\zeta$ suggests that the leading edge vortex advecting into the wake contributes to the shear layer corrugation present in both the perturbed model $\tilde{\ub}(\xb,t)+\ub'(\xb,t)$ (figures \ref{fig:osc-ftle-pert-a}$a,b$) and the original flow field (figure \ref{fig:osc-orig-ftle}). Notably, the $\zeta$ fields capture both the intensification of the vortex roll-up (see e.g., $x/c \approx 1.0-1.5$ and $y/c < 0$) and the corrugation of the shear layer. The instability of the LCS that mode $\ub_2$ induces may relate to the coherent Reynolds stresses observed from \citet{Jones_Kanso_Luhar_2024}, which correspond to thrust and high propulsive efficiency. These physical insights also complement prior work \citep[e.g.,][]{QuinnOpt,lentink_vortex-wake_2008} aiming to characterize the influence of local flow features around oscillating foils on the downstream wake structure. In general, these results show that the $\Delta_\infty ^2\mathrm{MS}$ and $\zeta$ fields serve as useful tools for relating Lagrangian transport effects to specific Eulerian flow features. 

\subsection{Example 3: Turbulent channel flow}
The previous sections have demonstrated the utility of the modal-trajectory uncertainty framework in the context of model reduction. Here, we use it to highlight the effect of specific modes on LCSs for a flow field with broadband length and time scales. 

\begin{figure}[t!]
    \centering
    \includegraphics[
        width=0.8\textwidth,
        keepaspectratio
    ]{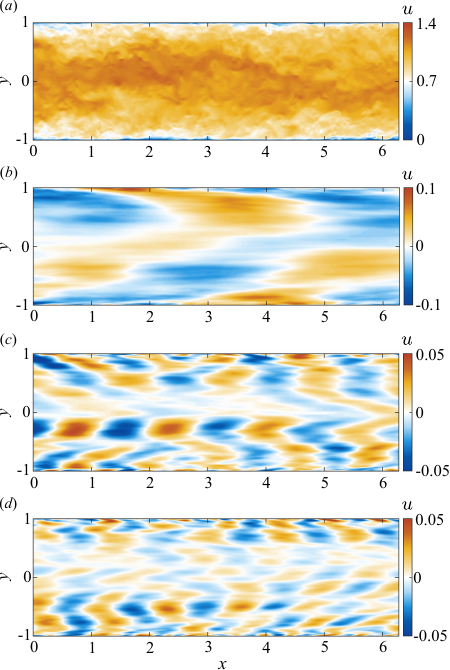}

    \caption{
    Streamwise velocity component of ($a$) the turbulent channel flow field at $t=0$, and the corresponding DMD modes with the following frequencies, wavelengths, and relative magnitudes:
    ($b$) $\Im(\omega)=1.1$, $\lambda_x/h\approx 6.0$,
    $|\ub'|/|\tilde{\ub}| \approx 0.026$;
    ($c$) $\Im(\omega)=4.0$, $\lambda_x/h\approx 1.5\text{--}2.0$,
    $|\ub'|/|\tilde{\ub}| \approx 0.015$;
    ($d$) $\Im(\omega)=5.1$, $\lambda_x/h\approx 1.0$,
    $|\ub'|/|\tilde{\ub}| \approx 0.012$.
    }

    \label{fig:gx-cturb}
\end{figure}

In particular, we use turbulent channel flow with a friction Reynolds number $Re_\tau = u_\tau h/ \nu =1000$ from the Johns Hopkins Turbulence Database \citep{graham_web_2016}, where $u_\tau$ is the friction velocity and $h$ is the channel half height. Two-dimensional data were extracted over the domain of $(x,y) \in [-\pi,\ 3\pi] \times [-1,\ 1]$ and yielded $1024 \times 160$ velocity vectors. Here, the dimensions are normalized by $h$. A set of $n_t=565$ snapshots were extracted with a time step of $\Delta t=0.02h/u_b$, where $u_b$ is the bulk velocity. Figure \ref{fig:gx-cturb}a shows the streamwise component of the velocity field for the first snapshot. We recognize that the use of a two-dimensional (2D) slice for LCS analyses is a substantial simplification as out-of-plane motion can significantly alter particle trajectories. However, prior work from \citep{green_detection_2007} has shown that this 2D approximation still provides substantial insight.

\begin{figure}[t!]
    \centering  \includegraphics[width=0.8\textwidth,keepaspectratio]{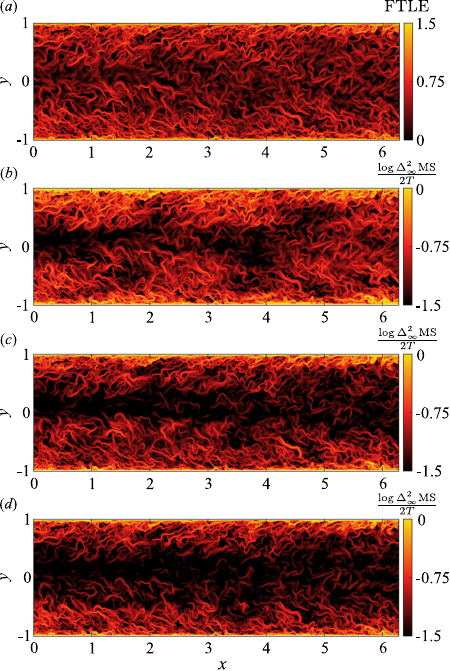}
    \caption{LCS structures for the turbulent channel flow ($a$) Forward FTLE field at $t=0$ and modal-trajectory uncertainties for large-scale structures: ($b$) $\Im(\omega)=1.1$, \ $\lambda_x/h\approx 6.0$, \ ($c$) $\Im(\omega)=4.0, \ \lambda/h\approx 1.5-2.0$, \ ($d$) $\Im(\omega)=5.1, \ \lambda_x/h\approx 1.0$.}
    \label{fig:cse-cturb}
\end{figure}

Tracer particles were advected for an integration time of $T=2.9$ in the subdomain of $(x,y) \in [0,\ 2\pi] \times [-1,\ 1]$. The initial grid comprised $514\times160$ particles. The selected subdomain allows for a majority of the particles to advect with the velocity field before exiting the domain at $x=3\pi$. Particles that exited the domain early continued to advect with the time-averaged flow field (as in equation \ref{eq:mean}) at the respective $y-$location. Small-scale filament structures are observed in the forward FTLE field of the original flow , as shown in figure \ref{fig:cse-cturb}$a$.

\begin{figure}[t!]
    \centering  \includegraphics[width=0.85\textwidth,keepaspectratio]{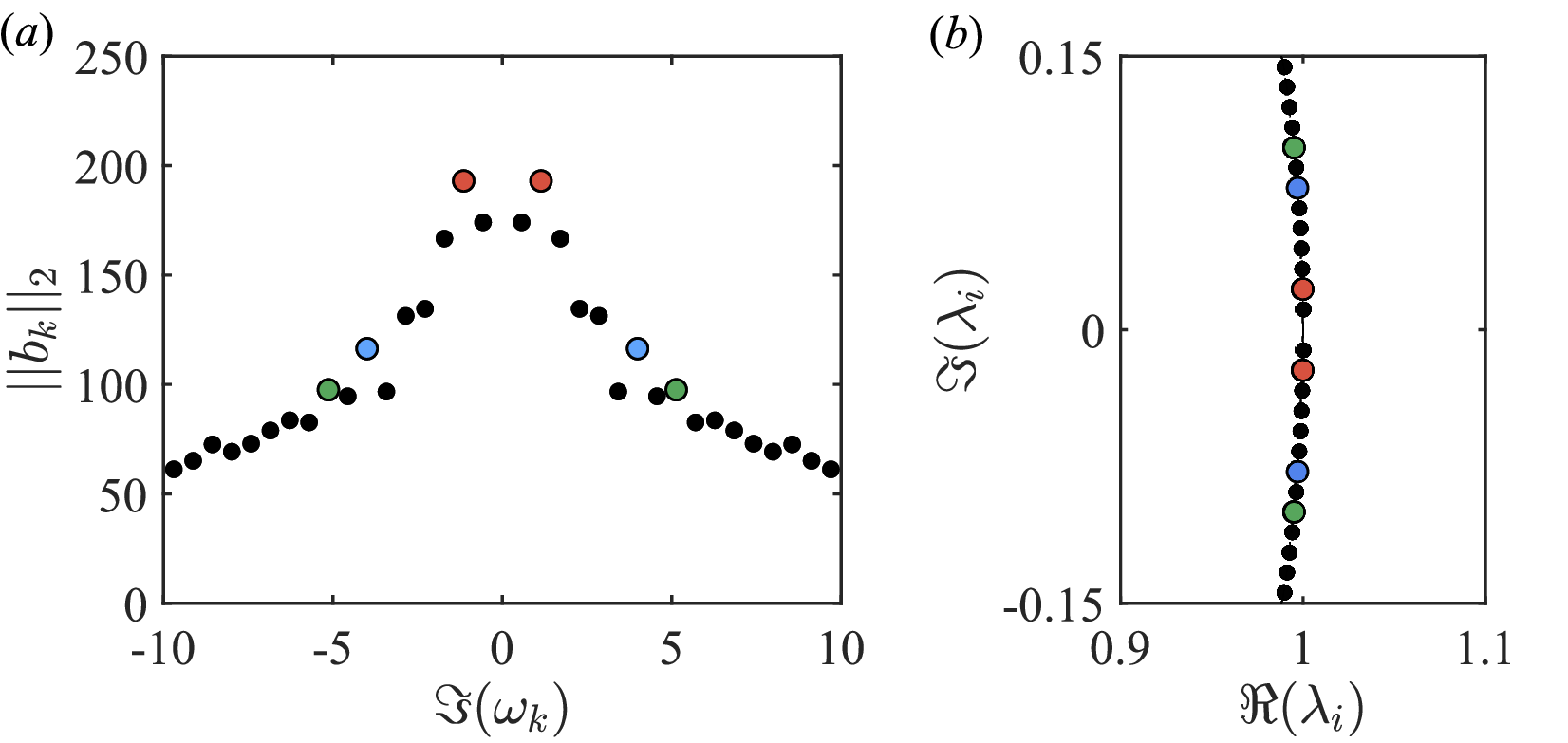}
    \caption{($a$) Amplitude and ($b$) eigenvalue spectrum of the DMD modes for the turbulent channel flow field. The mode pairs highlighted in red ($\Im(\omega)=1.1$), blue ($\Im(\omega)=4.0$), and green ($\Im(\omega)=5.1$) are separately chosen as the perturbations $\ub'(\xb,t)$.}
    \label{fig:spect-cturb}
\end{figure}
Previous studies have shown that modal analysis techniques such as SPOD and resolvent analysis can identify energetic modes with streamwise scales comparable to those associated with superstructures and other large-scale motions in wall-bounded turbulence \citep{towne_database_2023, saxton-fox_amplitude_2022}. The aim in this example is to use the modal-trajectory uncertainty framework to examine how such mode structures can influence the smaller scale filaments seen in figure \ref{fig:cse-cturb}$a$.

 For the present data, we use the DMD algorithm on the full domain to extract the dynamically coherent structures of interest. DMD was chosen because it associates each mode with a characteristic temporal frequency and growth/decay rate, which is useful for characterizing coherent structures in broadband turbulent flows containing multiple dynamically relevant frequencies. Because the channel-flow dataset is statistically stationary, we subtract the time-averaged flow $\bar{\bm{u}}$ before applying DMD; in the present case, this yields modes with decay rates close to zero, consistent with a Fourier-like representation of the fluctuations \citep{chen_variants_2012}.

As considered in the previous examples, we treat modes with the same oscillation frequency as pairs. Three local peaks in amplitude for each mode pair are highlighted in figure \ref{fig:spect-cturb} (see red, blue and green markers). The modes have streamwise wavelengths of $\lambda_x/h \approx 6.0, \ 1.5-2.0$, and $1.0$, where $h$ is the channel half-height, and are shown in figure \ref{fig:gx-cturb}. These streamwise lengths are similar to those of the large-scale motions found in prior studies \citep{kim_very_1999,monty_comparison_2009,smits_highreynolds_2011}; however, because the present analysis is two-dimensional, we do not identify the DMD modes themselves as LSMs or VLSMs. Instead, we interpret them as 2D modal approximations of energetic large-scale motions \citep{towne_database_2023}.

To better understand the interactions between these larger-scale features 
and smaller-scale filaments of the FTLE field, we consider 
three separate modal-trajectory uncertainty cases. Here, we define the full index set of $n_t-1$ modes as $\mathcal{H} = 
\{1, 2, \ldots, n_t-1\}$, which excludes the mean flow $\bar{\bm{u}}$, and for 
each case, let $\mathcal{G} \subset \mathcal{H}$ denote the selected 
subset of highlighted mode pairs treated as the perturbation $\ub'$. The 
complement $\mathcal{F}=\mathcal{H} \setminus \mathcal{G}$ then contains all mode 
indices \textit{except} those in $\mathcal{G}$. The baseline and 
perturbation fields then are defined as:
\begin{equation}
\tilde{\ub}(\xb,t)=
\bar{\ub}(\xb)+\sum_{k \in \mathcal{F}}
\ub_k (\xb,t),
\end{equation}
and
\begin{equation}
\ub'(\xb,t)=\sum_{k \in \mathcal{G}}
\ub_k (\xb,t),
\label{cturb-sys1}
\end{equation}
respectively, where $\ub_k(\xb,t)=b_k \exp(\omega_k t)\boldsymbol{\psi}_k$. We note that the previous examples used highly truncated models 
(rank $r \ll n_t$), which retain only the dominant modes and may filter out length scales important to turbulent flows.

Regions where the smaller scale structures are affected by these modes are shown in figure \ref{fig:cse-cturb}. The FTLE field for the full flow field in figure \ref{fig:cse-cturb}$a$ captures persisting ridges of sensitivity across the entire domain. In contrast, the modal-trajectory uncertainty $\Delta_\infty ^2\mathrm{MS}$ field only highlights the ridges that are influenced by the mode structure $\bm{u}_s$. For instance, the influence of the largest structure extends across the entire height of the channel (figure~\ref{fig:cse-cturb}$b$) while the influence of the smallest structure is greatest in the near-wall region (figure~\ref{fig:cse-cturb}$d$). Moreover, the $\Delta_\infty ^2\mathrm{MS}$ fields also show inclination in the $x-y$ plane that is consistent with mode structure. These observations show that modal-trajectory uncertainty can help identify flow structures associated with large-scale modes in turbulent flows.
\\ \\

\section{Discussion and Conclusions}

This paper applies the model sensitivity framework developed by \citet{kaszas_universal_2020} to characterize how LCSs are affected by specific modes identified using modal analysis techniques. The framework yields two important metrics: 1) the modal-trajectory uncertainty field $\Delta_\infty ^2\mathrm{MS}$ which highlights regions of the LCS structure in the fluid flow that a given mode $\ub'(\xb,t)$ is likely to affect, and 2) the FTLE perturbation $\zeta$ which provides more explicit details on where the mode perturbs the FTLE field. While these metrics are related to the FTLE, their patterns pertain more to the dynamic interactions between individual modes (or subsets of modes) and the LCS, adding a new perspective to modal decompositions. The examples in this paper suggest that the FTLE perturbation $\zeta$ is most useful for representing dynamics when applied to modal representations at scales larger than those of the perturbations (e.g., cylinder-wake and oscillating-foil examples), whereas the modal-trajectory uncertainty $\Delta_\infty ^2\mathrm{MS}$ is better suited for capturing large-scale structures acting on smaller-scale motions (e.g., turbulent-channel-flow example).

The method is demonstrated using both experimental and numerical data. These examples consider the relationship between LCSs and flow features such as cross-stream perturbations, vortex shedding, shear layer instabilities, and large-scale structures in turbulent flows. For these examples, and many other fluid flows that are not presented in this paper, the choice of the reconstructed flow field $\tilde{\ub}(\xb,t)$ (as defined in equation \ref{eq:CSeqn}) has considerable flexibility. It is not limited to any specific modal representation. In other words, the method can be extended to other decomposition methods that use different basis functions from the POD or DMD algorithms primarily used in this paper. The main requirement is that $\tilde{\ub}(\xb,t)$, and hence its FTLE field, contains a majority of the dynamics from the original flow field. The definition of the perturbation $\ub'(\xb,t)$ used in modal-trajectory uncertainty is also flexible. It could be a single mode or a collection of modes with different dynamic properties (e.g, frequencies, growth rates, topologies etc.).

An important aspect of the FTLE perturbation $\zeta$ is that it represents the Eulerian modes in a different feature space. Decomposition approaches derived from stability analyses often represent growing, decaying, or harmonic mode perturbations with respect to a mean flow. The FTLE perturbation represents perturbations to the FTLE flow structure of $\tilde{\ub}(\xb,t)$ due to the modes in $\ub'(\xb,t)$. In two of the examples considered in this paper, we use the DMD algorithm which identifies an oscillation frequency and growth rate for each mode. The corresponding FTLE perturbation may not preserve this interpretation. Instead, it represents DMD modes in a latent space based on the FTLE structure.

This framework can complement bispectral mode decomposition \citep{schmidt_bispectral_2020} and similar energy-transfer analyses, particularly in applications concerned with the spatial and temporal scales that govern mixing and transport. In stratified flows, for instance, vertical scalar transport is often suppressed while lateral mixing is enhanced through internal-wave dynamics \citep{riley_how_2022}. In this context, modal-trajectory uncertainty may help identify these transport barriers and how they evolve under different flow mechanisms, including internal waves with distinct morphologies and mixing or breaking characteristics \citep{chinta_regime_2022,ohh_wake_2022}.


This framework can also be expanded to consider mode truncations used to de-noise experimental data or filtered computational models such as large-eddy simulations. By definition, truncated modal representations or filtered numerical simulations only aim to capture the dominant dynamics. The $\mathrm{MS}$ framework could be used to characterize the influence of the neglected flow features on the FTLE fields, i.e., identify the uncertainty in LCS calculations pursued using filtered experimental or numerical data. Such analyses could be especially important in environmental or geophysical contexts where Lagrangian dynamics and transport play a vital role (e.g., for nutrient cycling, pollutant dispersal), whereas flow field estimates from numerical simulations, field measurements, or remote-sensed data are likely to be under-resolved.   

It is worth noting that the algorithm used to compute modal-trajectory uncertainty in this paper is computationally expensive when applied to three-dimensional flow fields. Approximate methods could be used to improve the speed by avoiding redundant computations of the flow map \citep[see e.g.,][]{brunton_fast_2010}. Nevertheless, the framework presented here could be used to better understand how modal representations obtained from experimental or simulation data affect Lagrangian dynamics and transport, i.e., to connect these Eulerian modes with Lagrangian coherent structures.
\\ \\
The data and code used for this work are available in a public
GitHub repository \citep{Jones_github}.

\section{Acknowledgments}
The authors would like to thank John O. Dabiri for insightful discussions related to this work and the two reviewers of this paper for their helpful feedback. Funding from the National Science Foundation under grant no. 1943105 is also gratefully acknowledged.
\begingroup
\RaggedRight
 \begin{spacing}{.89}
 \setlength{\bibsep}{5.5pt}
\bibliography{references}
 \end{spacing}
\endgroup
\end{document}